\newcommand{\D}{\Delta}

\newcommand{\la}{\lambda}
\newcommand{\La}{\Lambda}

\newcommand{\si}{\sigma}

\newcommand{\p}{\partial}%Differentiations

\newcommand{\non}{\nonumber}

\newcommand{\lb}{\left(}
\newcommand{\rb}{\right)}
\renewcommand{\le}{\left\lbrack}
\newcommand{\re}{\right\rbrack}

\documentstyle[12pt]{article}

\begin{document}

\def\beq{\begin{equation}}
\def\eeq{\end{equation}}
\def\jm{j^{\mu}}
\def\g5{{\gamma}_{5}}
\def\sm{$ {\rm T}_{\rm SM} $}
\def\beqa{\begin{eqnarray}}
\def\eeqa{\end{eqnarray}}
\def\b2{{\rm B}^{2}}
\def\mt{{\rm m}_{t}}
\def\m2{{\mu}^{2}}
\def\in{$\frac{1}{\rm N}$}
\def\fj{\frac{\partial f}{\partial m}}
\def\fd{\frac{\partial^2 f}{\partial m^2}}
\def\ft{\frac{\partial f}{\partial x}}
\def\fc{\frac{\partial^2 f}{\partial x^2}}

\newcommand{\etal}{\mbox{\it et al.}}
\newcommand{\prdj}[1]{{ \it Phys.~Rev.}~{\bf D{#1}}}
\newcommand{\prlj}[1]{{ \it Phys.~Rev.~Lett.}~{\bf {#1}}}
\newcommand{\plbj}[1]{{ \it Phys.~Lett.}~{\bf {#1B}}}
\newcommand{\npbj}[1]{{ \it Nucl.~Phys.}~{\bf B{#1}}}
\newcommand{\ptpj}[1]{{ \it Prog.~Theor.~Phys.}~{\bf {#1}}}
\newcommand{\prep}[1]{{ \it Phys.~Rep.}~{\bf {#1}}}

\titlepage
\begin{flushright} MPI-Ph/93-42\\
TUM-TH-157/93 \\
UW-IFT-10/93 \end{flushright}
%\vspace{0.7cm}

\begin{center} \bf

PHASE TRANSITIONS IN NJL AND SUPER-NJL MODELS${}^{*}$\\
\rm
\vspace{3ex}
Zygmunt Lalak${}^{a,b,1}$, Jacek Pawe\l  czyk${}^{a,1}$ and Stefan
Pokorski${}^{a,c,}${}\footnote[1]{lalak,pawelc,pokorski@fuw.edu.pl\\
${}^*$Work partially
supported by Polish Commite for Scientific Research.}~\\
\vspace{3ex}
${}^{a}$ Institute of Theoretical Physics \\ University of Warsaw \\
 Ho\.{z}a 69, Warsaw, PL 00-681 \\
\vspace{2ex}
${}^{b}$ Physics Department\\ Technische Universit\"{a}t
M\"{u}nchen\\ D-85748 Garching\\
\vspace{2ex}
${}^{c}$ Max-Planck-Institut f\"{u}r Physik\\
Werner-Heisenberg-Institut\\ P.O.Box 401212 M\"{u}nchen\\
\vspace{0.7cm}

ABSTRACT
\\
\end{center}

An elementary method of determination of the character of
the hot phase transition in 4d four-fermion NJL-type models is
applied to non-supersymmetric and supersymmetric versions
of simple NJL model. We find that in the non-susy case the
transition is usually of the second order.
It is weakly first order only in the region of parameters
which correspond to fermion masses comparable to the cut-off.
In the supersymmetric case both kinds of phase transitions
are possible. For sufficiently strong coupling and sufficiently
large susy-breaking scale
the transition is always of the first order.\\

\newpage

\noindent{\em 1. Introduction}.
Recently a lot of attention has been devoted to the problem of phase
transitions in
a variety of particle physics models. A sound motivation for this
activity is given by cosmology. In particular, it is believed that
first order phase transitions imply
several interesting and potentially observable effects. One well known
and widely considered example is the possibility of baryon number
generation during first-order electroweak phase transition [1].
 Other examples include
gravitational wave production during bubble collisions [2] and
modifications of relic dark matter particle abundancies [3].

As the symmetry breaking in the standard model is achieved by means of a scalar
field with a nonzero vacuum expectation value, it is
 the temperature phase structure of the bosonic models,
for example $\phi^4$ theory,  which
attracts at present attention of the public and is subject to a careful
investigation. At the same time similar problems in simplest fermionic models,
like the Nambu-Jona Lasinio (NJL) one, remain rather poorly known in four
dimensions.
It should be stressed that  the fermionic models, although generically
nonrenormalizable in 4d, are very useful  as phenomenological
effective Lagrangians. They facilitate quantitative studies of phenomena for
which fundamental theories are not known and provide a tractable description
of dynamical symmetry breaking. On more practical side, a specific interest in
 this type of theories  may be justified by
composite Higgs models of BHL-type [4],[5] and  by
recent revival of studies on
NJL-type QCD
condensates  [6].

In this note we shall discuss the temperature phase structure of the NJL and
super-NJL (SNJL) models to leading order in large N expansion.
An elementary  but
effective algorithm is used to determine  the type of hot phase transitions in
those models.
We study parameter space of the nonsupersymmetric and supersymmetric
four-fermion theories with particular regard to regions interesting for
 phenomenological applications.

The structure of the paper is as follows. In section 2 we describe
 our approach to hot phase transitions in NJL models, in section 3
we apply the algorithm to the nonsupersymmetric model which allows for a
fully analytical treatment, and in section 4 we elaborate on the more complex
supersymmetric  NJL case.

\vspace{1cm}
\noindent{\em  2. NJL models of fermion condensation at finite
temperature.}
We set out to investigate the 4d four-fermion models to the leading order in
large N expansion. Let us recapitulate briefly main steps of the standard
procedure which  converts the original Lagrangians into the
form suitable for $1/N$ expansion. The goal is achieved through the
 introduction of
auxiliary scalar fields [7],[8]. The original Lagrangian
 of the NJL model then reads:\footnote{We use notation of the book [13].}
\beqa
{L}_{NJL}&=&i\bar{\lambda} \gamma^{\mu}\p_{\mu} \lambda +\frac{g^2}{4}
[ (\bar{\lambda}\lambda)^2 - (\bar{\lambda} \gamma_{5} \lambda)^2 ]\non\\
&=&i\p_\mu\bar{\la}_+\bar{\si}^\mu\la_+ +
i\p_\mu{\la}_-{\si}^\mu\bar{\la}_- -B^*B+gB\la_-\la_++
gB^*\bar{\la}_+\bar{\la}_-
\label{eqj}
\eeqa%1
where $\la_+,\la_-$ are chiral Weyl components of $\la$ and $B$ is an
auxiliary scalar field, nonpropagating at tree level.
In order to apply the $1/N$ expansion to this lagrangian
one considers N generations
 of fermions
$\la$ and rescales $B\to \sqrt{N}B,\; g\to g/\sqrt{N}$. Integration
over fermions gives an effective action for the auxiliary field $B$.
The vacuum expectation value of that field is related to the fermion condesate
through the relation $<\bar{\lambda} \lambda > = \frac{2B}{g}$ valid
 to the leading order in $1/N$. Hence, the object of interest to us is the
 potential for $B$. To determine this potential it is sufficient to
consider real valued, homogeneous $B$.
 Denoting $m\equiv g B$ one gets the
effective action in the form
\beq
S_{NJL}[m]=N \; \int d^4 x\;\lb \frac{ m^2}{g^2} -2\, tr \;\log(-\p^2+m^2)\rb
\label{act}
\eeq%2
where  $S$ is written in the Euclidean space. We use finite
temperature imaginary time formalism due to Matsubara i.e. we compactify
 the euclidean time on a circle with circumference
 $\beta\equiv 1/T$.
The fermionic functional determinant in (\ref{act}) is calculated
with anti-periodic
boundary conditions in the compact direction.

The supersymmetric generalization of NJL model was constructed in [9].
The Lagrangian with auxiliary superfields has the form
\beqa
L_{SNJL}&=&\int d^4\theta [(\bar{\Phi}_+\Phi_+ +\bar{\Phi}_-\Phi_-)(1-
\D^2\; \theta\theta\bar{\theta}\bar{\theta}) + \bar{\Phi}_2\Phi_2]\non\\
&+&\le\int d^2\theta[\Phi_1\Phi_2-g\Phi_1\Phi_+\Phi_-]+c.c.\re.
\label{snjl}
\eeqa%3
All superfields in (\ref{snjl}) are chiral, $\Phi_{\pm}$
contain fermions $\la_+ ,\la_-$, $\Phi_{1,2}$ are auxiliary superfields and
$\D$ is a susy-breaking scale.
The first component of $\Phi_1$, which we denote by $A_1$, is
proportional to $<\bar{\la}\la>$ and $m \equiv g A_1$ is, as before,
the mass of the fermion $\la$.
Functional integration over all matter fields leads to the following
effective action for the parameter $m$
\beq
S_{SNJL}[m]=N\lb\int d^4x\,\frac{m^2}{g^2}-2\,tr \; \log(-\p^2+m^2)
+2\,tr\;\log(-\p^2+\D^2+m^2)\rb
\label{sact}
\eeq%4

Performing direct evaluation of
(\ref{act}) and (\ref{sact}) one obtains finite temperature effective
potentials $V_{NJL}$ and $V_{SNJL}$ respectively
\beqa
V_{NJL}(m,T)&=& \frac{m^2}{g^{2}} - \frac{1}{16 \pi^2 } \{ m^2 \Lambda^2 - m^4
\log(1+\frac{\Lambda^2}{m^2})\nonumber\\
& +& \Lambda^4 \log(1 + \frac{m^2}{\Lambda^2}) \}
- \frac{4}{\beta} \int \frac{d^{3} k}{(2 \pi)^3 } \log(1+e^{-\beta
\sqrt{m^2 +k^2}})
\label{dzs}
\eeqa%6
and
\beqa
V_{SNJL}(m,T,\Delta )&= &\frac{m^2}{g^{2}} - \frac{1}{16 \pi^2 } \{ m^2
 \Lambda^2
- m^4
\log(1+\frac{\Lambda^2}{m^2})\nonumber\\
& +& \Lambda^4 \log(1 +
\frac{m^2}{\Lambda^2})
-(m^2 + \Delta^2) [ \Lambda^2 \nonumber\\
&- &(m^2 +\Delta^2 ) \log(1+
\frac{\Lambda^2}{m^2 + \Delta^2})] - \Lambda^4 \log(1+ \frac{m^2 +\Delta^2
}{\Lambda^2}) \}\nonumber\\
&-&\frac{4}{\beta} \int \frac{d^{3} k}{(2 \pi)^3 } \log(1+e^{-\beta
\sqrt{m^2 +k^2}})\nonumber\\
& +& \frac{4}{\beta} \int \frac{d^{3} k}{(2 \pi)^3 }
\log(1-e^{-\beta
\sqrt{m^2 + \Delta^2 + k^2}})
\label{jj}
\eeqa%7
where the divergencies have been regularized  by a simple momentum
cut-off at the
scale $\La$.
 One should note at this point that all the  formulae we
have written till now are well defined for all values of parameters.
As discussed in the literature  ([10],[11],[12]) the form of $V_{NJL}$ and
$V_{SNJL}$
as functions of the temperature
determines the character of the phase transition between localized vacuum
configurations. Positions of the extrema of the potential are given by
the gap equation which is just the condition for vanishing of its first
derivatives with respect to $m$.

Our purpose is to determine the type of the phase transition as
a function of the parameters of the theory. First of all, let us make
the following
observation.
By direct inspection of eqs.(\ref{dzs}) and (\ref{jj}) one concludes
that the derivatives of both effective potentials with respect to $m$
consist of two parts: the first part is temperature independent, the second
temperature dependent. It is easy to check that the
temperature independent
 part is an increasing function of $m$  while the temperature dependent
one  is a decreasing function of $m$, with monotonically decreasing
(increasing) their first derivatives (i.e. the second derivatives of
the respective terms in the potential), respectively.
Moreover, the second derivatives of these functions (third derivatives
of the respective terms in the potential) never cross each other.
All together, this implies that each
 gap equation, which is the summ of the two terms, has at most
two  different from zero roots. The
greater of them always corresponds to a minimum because the potentials
 are increasing functions of $m$ for large enough $m$.
Hence these effective potentials
have at most one nontrivial minimum, at any temperature.
% Having determined this we can
%actually find the order of the phase transition looking at the vicinity of the
%point $m=0$ (this will be clarified soon).

There is another observation which simplifies the reasoning -the effective
potentials of both models under consideration are analytic
functions of $m^2$ around $m=0$ for nonzero $T$ and $\Delta$.
This is due to the additional mass gap which appears for fermions
compactified on a circle with antiperiodic boundary conditions.
Hence one can expand the finite temperature effective potentials in the
vicinity of $m=0$ in
the Taylor series in the variable $m^2$.

We define the critical temperature $T_c$ in the standard way,
demanding
$\frac{\p^2 V(T_c)}{\p m^2}|_{m=0} =0 =  \frac{\p V(T_c)}{\p (m^2)}|_{m=0}$.
Of course, below $T_c$ the $\frac{\p V}{\p m^2} |_{m=0}$ is negative
and above $T_c$ -- positive.
At this critical temperature the term in the expansion which is
proportional to $m^4$ can be negative or positive.\footnote{We disregard
the case when it vanishes as the one which requires a fine tuning of
parameters.}
In the former case, at $T=T_c$ the potential has to have a
minimum for $m\neq 0$. The phase transition which may take place at
$T \neq T_c$ is of the first order.
This follows from the fact that if one slightly increases the temperature
above $T_c$, one gets $\frac{\p V}{\p (m^2)}|_{m=0}>0$ i.e.
a "hill" arises between the local minimum at $m=0$ and the global
 minimum which
exists for $m>0$ due to $\frac{\p^2 V}{\p (m^2)^2}|_{m=0,T=T_c}<0$.
In the second case the phase transition takes place at $T=T_c$ and
is of the second order. Indeed, for $T > T_c$ the potential has a minimum
at $m=0$ and
 slightly
below $T_c$ one has $\frac{\p V}{\p (m^2)}|_{m=0}<0$ which implies that
a "valley" of an infinitesimal depth  arises  next to $m=0$.
There cannot be any other
non trivial minimum for $m\neq 0$ as this would contradict
the observation that the potential can have only one
nontrivial minimum.

In conclusion, we can determine the order of the phase transition
by studying the behaviour of the effective potential in the vicinity of
the point $m=0$.

In the next two sections we shall analyse phase transitions
in NJL and SNJL models according to the prescription presented in this
section.

\vspace{1cm}
\noindent{\em 3. Nonsupersymmetric NJL model at finite temperature.}
Let us warm-up with the simpler, nonsupersymmetric model. Since we are
operating in the vicinity of $m=0$ and the effective potential is an
analytic function at that point we can Taylor  expand it
in $m^2$, as explained in section 2.
\beqa
V_{NJL}(m,T)&=& V_{NJL}(m=0,T) + m^2 (\frac{1}{12 \beta^2} +
              \frac{1}{g^{2}} - \frac{\Lambda^2}{8 \pi^2})\nonumber\\
&+ &  m^4 \frac{1 + 2 c_f + 2 \log (\beta^2 \Lambda^2)}{32 \pi^2} +
       o(m^6 \beta^2)
\label{jt}
\eeqa%9
where $\beta=1/T$, $c_{f}=2 \gamma -3/2 -2 \log \pi \approx -2.64$.

According to the program outlined in section 2, we find the critical
temperature $T_c$ demanding that the $m^2$ term in (\ref{jt})
vanishes at $T_c$. We get
\beq
T_c^2 = \frac{3 \Lambda^2}{2 \pi^2} (1 -c)
\eeq%10
where $c=\frac{8 \pi^2}{g^2 \Lambda^2}$.

Next, the order of the phase transition is determined by the sign of the $m^4$
term at $T=T_c$.
The analysis of the different physical situations is straightforward
thus we simply summarize the results.
For $c>1$ - there is no phase transition as there is no (chiral)
symmetry breaking at $T=0$.
For $ 1 - \frac{2 \pi^2}{3} e^{c_{f} + 0.5}<c<1$ -
the phase transition is of the second order and it take place
at $T_c$. For
$0<c< 1 - \frac{2 \pi^2}{3} e^{c_{f}+0.5} \approx 0.22$ -
the phase transition is of the first order and it may take place
at $T>T_c$. One should note that the squared temperature of the
phase transition is greater then $\La^2 e^{c_{f}+0.5}
\approx \La^2/10$ in this case.

Looking at the $T=0$ gap equation one can see that the last
scenario corresponds to fermion masses comparable
to the cut-off $\Lambda$. This may be relevant for QCD, [6], but it is
 definitely not the range
of fermion masses suitable for top-mode standard model of BHL [4],
where to have the top mass of the order of $10^2$ GeV one needs
$\frac{8 \pi^2}{g^2 \Lambda^2} \approx 1 - O(10^{-26})$. The top-mode
scenario falls then safely into the region of the second-order phase
transition (to the leading order in large N), which confirms earlier
results [14].

\vspace{1cm}
\noindent{\em 4. Susy-NJL model at finite temperature.}
The construction of the supersymmetric extension of the simplest
version of the NJL model has been discussed carefully in [9]. It turns out
that in exactly supersymmetric theory fermion condensation is impossible
for any choice of parameters. However, if one introduces a soft susy-breaking
scalar mass $\Delta$, then the fermion condensate may appear and chiral
symmetry breaking becomes possible. Following the same procedure as in the
non-susy case we obtain to the leading order in large N, Taylor expanding
in $m^2$, the supersymmetric counterpart  of
(\ref{jt})
\beqa
V_{SNJL}(m,T,\Delta)&-&V_{SNJL}(0,T,\Delta)=\nonumber\\
&+& m^2 ( \frac{1}{12 \beta^2} +
              \frac{1}{g^{2}} - \frac{\Delta^2}{8 \pi^2} \log(1+
\Lambda^2 / \Delta^2) +
    \frac{1}{\beta^2 \pi^2} I_{2} (\beta^2 \Delta^2))\nonumber\\
&+& m^4 (\frac{1+2 c_f}{32 \pi^2} +
    \frac{\Lambda^2}{16 \pi^2 (\Lambda^2 + \Delta^2)}
\nonumber\\
&+& \frac{1}{16 \pi^2} \log (\frac{\beta^2 \Lambda^2 \Delta^2}{\Lambda^2
+\Delta^2})
    - \frac{1}{4 \pi^2} I_{0} (\beta^2 \Delta^2))
    + o(m^6 \beta^2)
\label{fs}
\eeqa%9
where  $I_{n}(x)=\int_{0}^{\infty} dy \, y^n (y^2 + x)^{-1/2} /
(\exp (\sqrt{y^2 +x} -1)$.\\
Following the procedure applied already in section 3 we find the
equation for the critical temperature demanding that
the $m^2$ term in formula (\ref{fs}) vanishes:
\beq
 \frac{1}{12 \beta^2} +
              \frac{1}{g^{2}} - \frac{\Delta^2}{8 \pi^2} \log(1+
\Lambda^2 / \Delta^2) +
    \frac{1}{\beta^2 \pi^2} I_{2} (\beta^2 \Delta^2)=0
\label{tc}
\eeq%10
The resulting equation cannot be
solved algebraically due to the presence of the function $I_2$.
In order to give a flavour of the dependence of $T_c$ on
parameters of the theory  we can find a rough
 approximate solution noting that values of the function $I_2$ range
from $\pi^2/6$ to $0$ at $\infty$. We arrive at the formula
\beq
\frac{T_c^2}{\Lambda^2}\approx  \frac{3}{2 \pi^2} \lb\frac{\Delta^2}{\Lambda^2}
\log(\frac{\La^2}{\D^2}+1) - c\rb.
\label{app}
\eeq%11

We  also see that for $\D<<\La$ the
term proportional to $m^4$ in eq.(\ref{fs})
depends only on
the product $\beta\D$. The condition for vanishing of this term can be
easily solved. We get that if $T_c<\D/2.16$  the phase transition
is of the second order. For $T_c>\D/2.16$ the phase transition is of
the first order. The characteristic scale which determins
the critical temperature is now the susy-breaking scale.

The analytic results one can get using the above
approximations can be summarized as follows.
If $c>1$ - there cannot be any phase transition -- the symmetry
remains unbroken.
If $0.07<c<1$ - for small susy-breaking scale, $0<\D<\D_1$,
again no phase transition is possible; for $\D>\D_1$ there is second order
phase transition. Its temperature is given by eq.(\ref{app}). The
boundary value of $\D$ i.e. $\D_1$ is given by
$\frac{\p V_{SNJL}}{\p m^2}(T=0,\D_1)=0$
(eq.(\ref{tc})).
If $\;0<c<0.07$ - for  $0<\D<\D_1$
no phase transition is possible ($\D_1$ is as above); for
$\D_1<\D<\D_2$ there is a second order phase transition; for
$\D>\D_2$ there is a first order phase transition. The value of
$\D_2$ is given by the condition that both terms on the r.h.s. of
(\ref{fs}) vanish.

The exact results of the numerical study are presented in the figure 1.
All values  in this figure are scaled by the cut-off $\La$.
The curve $``d1''$ represents the dependence of $d1=\D_1/\La$ on $c$
($``d1''$ is
the solution to the eq.(\ref{tc}) at  $ T=0$).
The curve $``d2''$ represents $d2\equiv \D_2/\La$.
Thus, above the curve $``d1''$ no phase transition is possible,
between $``d1''$ and $``d2''$ only the second order phase
transition occurs, below $``d2''$ the first order phase transition occurs.
The remaining curves are solutions to the equation (10)
for $T/\La=0.1,\; 0,2,\; 0.25$ respectively. These are contours of the fixed
critical temperature in the plane (d,c).

The
phase transition in the super-top-mode scenario, [5], described
 in the leading
order of the large N expansion of the SNJL model, can be of the first
order as long as one  expects that the critical temperature
for the electroweak breaking is close to $200 \;{} $
GeV and the supersymmetry
 breaking
scale $\Delta$ lies in the TeV range. There $T_c > \Delta / 2.16$
and in addition for fermion mass $O(200) \;{}$GeV we get $\Delta_2
 < \Delta$.

\vspace{1cm}
\noindent{\em 5. Summary.}
In summary,  a simple
method has been used to  determine  the character of
the hot phase transition in 4d four-fermion NJL model, both in its
 non-supersymmetric and supersymmetric version.
We  conclude that in the non-susy case the
transition may be of the first order, however in the region of parameters
which correspond to fermion masses comparable to the cut-off.
In the supersymmetric case also both kinds of phase transitions
are possible. For sufficiently strong coupling and
sufficiently large susy-breaking scale
the transition is always of the first order. We applied these results
 to models under current investigation. We have found that
the super-NJL scenario of the top quark
condensation can exhibit the first order phase transition in the physically
interesting range of parameters. \\
\\
Authors thank J. Wosiek and K. Zalewski for useful discussion.\\
Z.L. was supported by A. von Humboldt Foundation, also partially
supported by Deutsche
Forschungsgemeinschaft and EC grant SC1-CT91-0729.
\\
\\
\vspace{1cm}
\newpage
\noindent{\large References }
\vspace{2ex}

\begin{description}

\item{[1]}
V.Kuzmin,V.Rubakov,M.Shaposhnikov, \plbj{155}, 36 (1985).
\item{[2]}
A.Kosowsky,M.Turner, Fermilab-Pub-92-295-A, (1992).
\item{[3]}
S.Dimopoulos,R.Esmailzadeh,L.Hall,N.Tetradis, \plbj{247}, 601 (1990).
\item{[4]}
W.Bardeen,C.Hill,M.Lindner, \prdj{41}, 1647 (1990).
\item{[5]}
T.Clark,S.Love,W.Bardeen, \plbj{237}, 235 (1990), M.Carena, T.Clark,
C.E.M.Wagner,W.Bardeen,K.Sasaki, \npbj{369}, 33 (1992).
\item{[6]}
G.Ripka, "Introduction to Nambu Jona-Lasinio Models Applied to
Low Energy Hadronic Matter", in {\it Hadrons and Hadronic Matter},
eds. D.Vautherin, F.Lenz and J.W.Negele, Plenum Press, New York, 1990;\\
M.Bander,H.Rubinstein, \plbj{289}, 385 (1992).
\item{[7]}
A.Polyakov, {\it Gauge Fields and Strings}, Harwood Academic Publishers,
1987.
\item{[8]}
Z.Lalak, \plbj{278}, 284 (1992).
\item{[9]}
W.Buchmuller,U.Ellwanger, \npbj{245}, 237 (1984).
\item{[10]}
M.Sher, \prep{179}, 273 (1989).
\item{[11]}
R.Rivers, {\it Path Integral Methods in Quantum Field Theory},
Cambridge University Press, 1987.
\item{[12]}
L.Dolan,R.Jackiw, \prdj{9}, 3320 (1974).
\item{[13]}
J.Wess and J.Bagger, {\it Supersymmetry and Supegravity}, Princeton
University Press, Princeton, NJ, 1983.
\item{[14]}
S.Kawati,H.Miyata, \prdj{23}, 3010 (1981).\\
T.Hatsuda, T.Kunihiro, \plbj{198}, 126 (1987).\\
S.Klimt,M.Lutz,W.Weise, \plbj{249}, 386 (1990).
\end{description}
\newpage
\noindent{\large Figure caption}\\
{}\\
{}\\
Figure 1. Exact results of the numerical study of the expression (9).
Label d denotes the ratio $\Delta / \Lambda$, c represents
$\frac{8 \pi^2}{g^2 \Lambda^2}$.
The curve $``d1''$ represents dependence of $d1=\D_1/\La$ on $c$ being
the solution to the eq.(\ref{tc}) at  $ T=0$.
The curve $``d2''$ represents $d2\equiv \D_2/\La$.
The remaining curves are solutions of the equation (10)
for $T/\La=0.1,\; 0,2,\; 0.25$ respectively. These are contours of the fixed
critical temperature in the plane (d,c).

\end{document}